\documentclass[conference]{IEEEtran}
\IEEEoverridecommandlockouts
\usepackage{cite}
\usepackage[normalem]{ulem}
\usepackage{amsmath,amssymb,amsfonts,bbm}
\usepackage{algorithmic} 
\usepackage{graphicx}
\usepackage{textcomp}
\usepackage{xcolor}
\usepackage{float}
\usepackage[caption=false,font=footnotesize]{subfig}
\def\BibTeX{{\rm B\kern-.05em{\sc i\kern-.025em b}\kern-.08em
    T\kern-.1667em\lower.7ex\hbox{E}\kern-.125emX}}

\usepackage{pgfplots}
\pgfplotsset{compat=1.18}

\begin{document}

\title{Point Cloud Transmission \\ With Patch-Based Rate Adaptation\\
\thanks{This work was partially supported by the Wallenberg AI, Autonomous Systems, and Software Program (WASP) funded by the Knut and Alice Wallenberg Foundation, and by the Excellence Center at Linköping - Lund in Information Technology (ELLIIT). Training and inference used resources provided by the National Academic Infrastructure for Supercomputing in Sweden (NAISS), partially funded by the Swedish Research Council through grant agreement no. 2022-06725.}
}

\author{
\IEEEauthorblockN{
Shipeng Liu, Garima Bazard, Khac-Hoang Ngo, Per-Erik Forssén
}
\IEEEauthorblockA{
Department of Electrical Engineering, Linköping University, Linköping 58183, Sweden \\
Email: \{shipeng.liu, garima.bazard, khac-hoang.ngo, per-erik.forssen\}@liu.se
}
}

\maketitle

\begin{abstract}
We propose a joint source-channel coding (JSCC) framework for efficient wireless point cloud transmission. In the proposed scheme, the point cloud is first divided into local patches and transformed into compact latent representations, which are then mapped to symbols for transmission over a noisy channel.
Our model incorporates an entropy model that exploits inter-patch spatial dependencies to estimate patch-level coding costs and allocate different numbers of symbols accordingly. Unlike existing patch-based point cloud JSCC methods that use the same transmission rate between patches, we adapt the channel usage of each patch latent representation. To exploit inter-patch dependencies without patch-by-patch autoregressive processing, we select a group of anchor patches whose features provide contextual information for entropy estimation of non-anchor ones. 
The framework is trained using a reconstruction loss and an entropy-based rate surrogate. Experiments on ModelNet40 demonstrate overall rate–distortion improvements over existing baselines across channel conditions.
\end{abstract}


\section{Introduction}
\label{sec:intro}

Rapid development of autonomous systems, extended reality, robotic perception, and digital twins has made three-dimensional (3D) point clouds an important data modality for both machine and human perception. Unlike
images and videos defined on regular grids, point clouds represent 3D geometry as unordered and sparse spatial samples, creating substantial challenges for real-time wireless transmission due to their large data volume, irregular structure, and sensitivity to geometric distortion.

Conventional wireless point cloud transmission typically follows a separate source-channel coding (SSCC) pipeline, where point clouds are first compressed using codecs such as Geometry-based Point Cloud Compression (G-PCC)~\cite{graziosi2020overview} and deep learning-based point cloud compression (DPCC) scheme, such as~\cite{zhang2022transformer}.
Such systems suffer from both \emph{cliff and leveling effects}: the reconstruction quality degrades abruptly when the channel condition falls below the designed operating point, while it saturates once reliable decoding is achieved and therefore cannot benefit from further improvement in channel quality~\cite{gunduz2025joint}. Joint
source-channel coding (JSCC) instead maps source features directly to transmitted symbols and provides graceful degradation under channel variations~\cite{gunduz2025joint}.

Existing point cloud JSCC methods have explored voxelized or octree-based representations~\cite{lyu_spawc2025,deepjscc_pcg}, sparse-convolutional architectures~\cite{deep_jscc}, implicit neural representations~\cite{ota_learning_based}, generative decoding~\cite{yang2026channel}, and multi-level semantic representation~\cite{ying2025joint}.
Furthermore,~\cite{patch-based_pc_compression} proposed a patch-based autoencoder for point cloud geometry compression, which was later adapted in~\cite{semcom} for point cloud JSCC. This patch-based approach preserves fine-grained local geometry. However, the scheme in~\cite{semcom} maps all patches to equal-length symbol sequences, without accounting for variations in local geometry.

Entropy-guided variable-rate JSCC provides a mechanism for adaptive channel allocation by assigning symbols according to the estimated information content of latent features~\cite{balle2018,ntscc,deep_jscc}.
However, accurate entropy estimation for patch-based point clouds remains difficult because local patches are irregularly distributed and strongly correlated. Surface continuity and overlapping neighborhoods cause adjacent patches to contain partially redundant geometry. Ignoring this dependency leads to inefficient channel allocation.

To address this problem, we propose a JSCC framework, called GeoContext-JSCC, for efficient patch-based point cloud transmission. 
An entropy model exploits inter-patch spatial context to estimate patch-level coding costs and allocate different numbers of symbols accordingly.
To exploit inter-patch dependencies, we divide the patches into anchor and non-anchor groups. The anchor patches are processed first, and their features provide shared context for non-anchor entropy estimation, encoding, and decoding. The patch centroids are transmitted over a digital link and combined with the recovered latent representations for point cloud reconstruction.

Experiments on ModelNet40~\cite{wu15_modelnet} demonstrate consistent rate--distortion improvements over SSCC schemes with G-PCC~\cite{graziosi2020overview} and DPCC~\cite{zhang2022transformer}, as well as  existing point cloud JSCC methods~\cite{wireless_point_cloud_transmission} and the fixed-rate variant across multiple transmission rates and channel conditions~\cite{semcom}. Ablation results further show that both entropy-guided channel-use allocation and inter-patch contextual coding in our design contribute to the reconstruction performance, and the proposed framework maintains graceful degradation as the channel quality decreases, as opposed to G-PCC and DPCC-based methods.

\textit{Notation:} 
We denote scalars by italic letters, vectors by bold lowercase letters, matrices and tensors by bold uppercase letters, and sets by calligraphic letters. We use $\mathbb R$ and $\mathbb C$ to denote the sets of real and complex numbers, respectively. For a positive integer $n$, let $[n]=\{1,\ldots,n\}$.

\section{System Model and Problem Formulation}

We consider a point-to-point wireless transmission system for 3D point clouds. Let 
$\mathcal{P}=\{\mathbf p_n\}_{n=1}^{N}$ denote an input point cloud containing $N$ points $\mathbf p_n\in\mathbb{R}^{3}$ to be transmitted from a transmitter to a receiver. The transmitter uses an encoder $f$ to map $\mathcal{P}$ to a sequence of symbols $\mathbf{x}\in\mathbb{C}^{N_{\mathrm{c}}(\mathcal{P})}$, and transmits it over~$N_{\mathrm{c}}(\mathcal{P})$ complex channel uses. The transmitted symbols satisfy the average power constraint
$
\lVert\mathbf{x}\rVert_2^2/N_{\mathrm{c}}(\mathcal{P})\leq P.
$ 
As in~\cite{lyu_spawc2025,deep_jscc},
we measure the transmission rate using the channel bandwidth ratio
(CBR), defined as
\begin{equation}
\mathrm{CBR}(\mathcal P)
=
\frac{N_c(\mathcal P)}{3N}.
\end{equation}

The receiver observes
\begin{equation}
    \mathbf{r}=\mathbf{x}+\mathbf{w},
\end{equation}
where~$\mathbf{w}\sim\mathcal{CN}
(\mathbf{0},\sigma^2\mathbf{I})$ denotes the additive white Gaussian noise (AWGN). This model is obtained after channel equalization has compensated for the fading effect. The signal-to-noise
ratio~(SNR) is defined as $\mathrm{SNR} = 10\log_{10} (P/\sigma^2) \, \mathrm{dB}.$
The receiver uses a decoder $g$ to map $\mathbf{r}$ to a
reconstructed point cloud $\widehat{\mathcal{P}}=\{\widehat{\mathbf p}_n\}_{n=1}^{\widehat N}$ containing $\widehat N$ points $\widehat{\mathbf p}_n\in\mathbb{R}^{3}$.

We optimize the encoder and decoder to minimize the rate--distortion objective:
\begin{equation}
\min_{f,g}
\mathbb{E}
\left[
\mathcal D(\mathcal P,\widehat{\mathcal{P}})
+
\lambda\,\mathrm{CBR}(\mathcal P)
\right], \label{RD_lagrangian}
\end{equation}
where the expectation is taken over the input point clouds and
channel noise, $\mathcal D(\mathcal P,\widehat{\mathcal{P}})$ denotes
the reconstruction distortion, and $\lambda>0$ controls the
trade-off between rate and distortion.
We choose $\mathcal D(\mathcal P,\widehat{\mathcal{P}})$ as the Chamfer distance~\cite{fan2017point}, defined below.

Consider two point clouds $\mathcal P=\{\mathbf p_i\}_{i=1}^{N}$ and
$\widehat{\mathcal{P}}=\{\widehat{\mathbf p}_j\}_{j=1}^{\widehat{N}}$. The nearest-neighbor indices are defined as~$j^*(i)=\arg\min_{j \in [\widehat{N}]} \|\mathbf p_i-\widehat{\mathbf p}_j\|_2^2$ and~$i^*(j)=\arg\min_{i \in [N]} \|\widehat{\mathbf p}_j-\mathbf p_i\|_2^2$, respectively.
The directed point-to-point errors are defined as
\begin{equation}
\label{eq:d1_directed}
d_{\mathrm{D1}}(\mathcal P,\widehat{\mathcal{P}}) =
\frac{1}{|\mathcal P|}
\sum_{i:\mathbf p_i\in\mathcal P}
\left\|
\mathbf p_i-\widehat{\mathbf p}_{j^*(i)}
\right\|_2^2.
\end{equation}

The symmetric Chamfer distance is defined as
\begin{equation}
    \mathcal D_{\mathrm{CD}}(\mathcal P,\widehat{\mathcal{P}})
    = d_{\mathrm{D1}}(\mathcal P,\widehat{\mathcal{P}})+d_{\mathrm{D1}}(\widehat{\mathcal{P}},\mathcal P).
\label{eq:chamfer_distance}
\end{equation}

Denoting the normals at $\mathbf p_i$ and
$\widehat{\mathbf p}_j$ of the corresponding point cloud by $\mathbf n_i$ and $\widehat{\mathbf n}_j$,
respectively, we define the directed point-to-plane errors as
\begin{equation}
d_{\mathrm{D2}}(\mathcal{P}, \widehat{\mathcal{P}})
=\frac{1}{N}\sum_{i=1}^{N}
  \left[
  (\mathbf p_i-\widehat{\mathbf p}_{j^*(i)})^{\mathsf T}
  \widehat{\mathbf n}_{j^*(i)}
  \right]^2.
\end{equation}

The peak SNR (PSNR) for the point cloud is defined as
\begin{equation}
    \mathrm{PSNR}_{\mathrm{D}q}
    =10\log_{10}
    \Bigg(\frac{6}{d_{\mathrm{D}q}(\mathcal{P}, \widehat{\mathcal{P}})+d_{\mathrm{D}q}( \widehat{\mathcal{P}},\mathcal{P})}\Bigg) \, \mathrm{dB} 
\end{equation}
for $q\in\{1,2\}$. We also write $\mathrm{PSNR}_{\mathrm{D}q}$ as $\mathrm{D}q$ PSNR.

\section{Proposed Scheme}

\begin{figure*}[t]
    \centering
    \includegraphics[width=\textwidth]{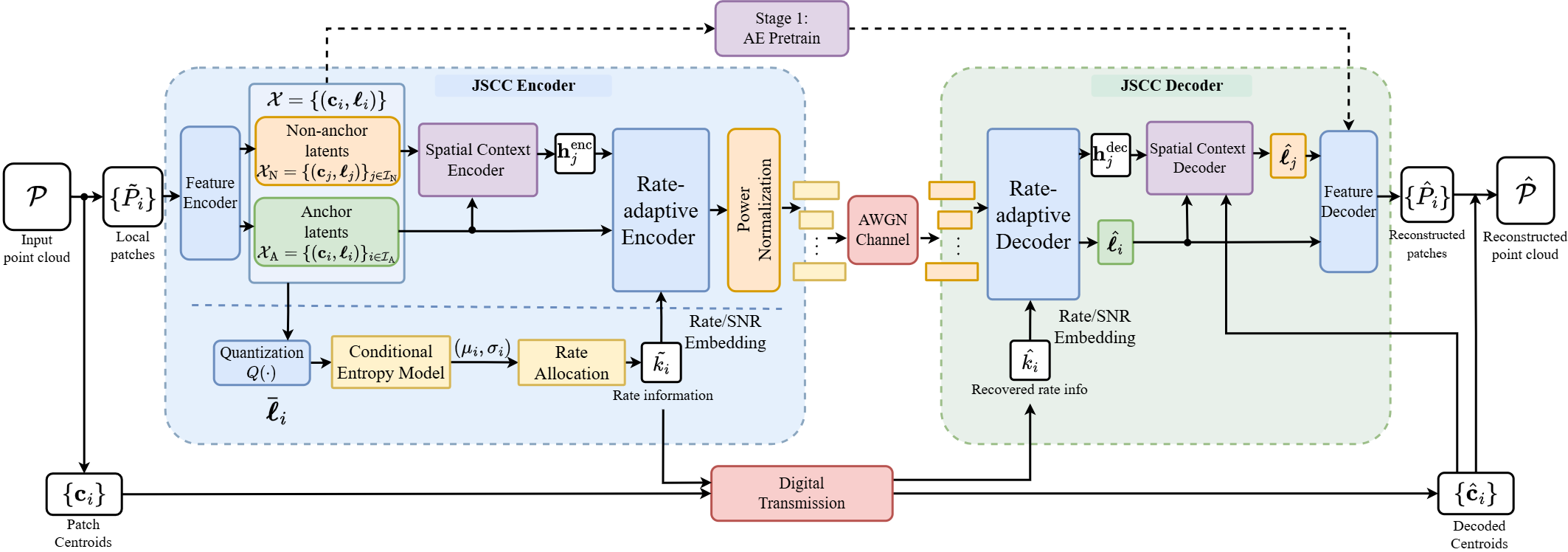}
    \caption{Architecture of the proposed GeoContext-JSCC framework for point cloud transmission with patch-based rate adaptation. The point cloud is encoded into patch representation $\mathcal{X} = \{  (\mathbf c_i, \boldsymbol{\ell}_i) \}_{i=1}^S$. Anchor patches provide contextual information for non-anchor encoding, while reconstructed anchor features support non-anchor decoding.
    The conditional entropy model exploits anchor latents and patch centroids to estimate patch-wise coding costs and determine rate assignments.
    Patch centroids and rate indicators are transmitted via digital communication. The dashed bypass is used for Stage~1, autoencoder pretraining.}
    \label{fig:system_overview}
\end{figure*}

We propose a hybrid analog--digital transmission framework, illustrated in Fig.~\ref{fig:system_overview}, and refer to it as \emph{GeoContext-JSCC}.
In this framework, we first partition the point cloud into local patches, each represented by a centroid and a latent vector. The patch latents are JSCC-encoded into symbol sequences with variable lengths determined by their estimated (conditional) entropy. These symbols are transmitted in an analog manner. Meanwhile, the patch centroids and the rate allocation indicator are transmitted using standard digital communication. To represent dependencies between patches while keeping the entropy model tractable, we select a group of anchor patches whose features provide contextual information for non-anchor ones.
At the receiver, the decoded local patch geometries are translated by the recovered centroids and assembled to reconstruct the point cloud. We next describe in detail the modules of this framework.

\subsection{Patch Construction and Representation}
\label{subsec:patch_construction}

We partition the input point cloud
$
\mathcal{P}=\{\mathbf p_n\}_{n=1}^{N}, 
\mathbf p_n\in\mathbb{R}^{3},
$
into $S$ 
local patches. Specifically, we employ farthest point sampling (FPS)~\cite{fps} to sequentially select $S$ representative points that are well separated.
We then group the $i$-th selected point and its $K-1$ nearest neighbors, identified using K-nearest-neighbor (KNN) search, to form local patch~$\mathcal{P}_i$, $i \in [S]$. The patches can overlap, and we call $\rho = KS/N$ the oversampling factor.
Let
$
\mathbf c_i
=
\frac{1}{K}
\sum_{\mathbf p\in\mathcal{P}_i}\mathbf p
$
denote the centroid of the $i$-th patch.
We express the patch points 
in local coordinates relative to~$\mathbf c_i$, yielding
$
\widetilde{\mathcal{P}}_i
=
\{\mathbf p-\mathbf c_i\colon \mathbf p\in\mathcal{P}_i\}.
$
Each centered patch is then mapped by a feature encoder
$f_{\mathrm{p}}$ to a latent representation
$
\boldsymbol{\ell}_i
=
f_{\mathrm{p}}(\widetilde{\mathcal{P}}_i) \in\mathbb{R}^{D}.
$
The feature encoder adopts the pre-quantization analysis transform
of~\cite{patch-based_pc_compression}, which consists of a per-point
set-abstraction block followed by a PointNet layer.
We represent the resulting collection of patch representations as
$
\mathcal{X}
=
\left\{
(\mathbf c_i,\boldsymbol{\ell}_i)
\right\}_{i=1}^{S}.
$

\subsection{JSCC Transmission of Patch Latents}
\label{subsec:anchor_context}

\subsubsection{Anchor-Conditioned Context}

We designate $S_{\mathrm A}$ patches as anchors to provide spatial context for the remaining $S-S_{\mathrm A}$ non-anchor patches. Specifically, we choose the first $S_{\mathrm A}$ patches selected by FPS as anchors to form a spatially distributed basis.\footnote{We observe empirically that 
the cosine similarity between anchor and non-anchor latents decreases with centroid distance, which motivates the use of nearby anchors to model local context of non-anchor patches.}
Index the anchor and non-anchor patches, respectively, by~ $\mathcal{I}_{\mathrm{A}}=[S_{\mathrm{A}}]$ and~$\mathcal{I}_{\mathrm{N}}=\{S_{\mathrm{A}}+1,\ldots,S\}$.
Accordingly, we partition the set of patch representations $\mathcal X$ introduced above into
~$
\mathcal X_{\mathrm A}
=
\left\{
(\mathbf c_i,\boldsymbol{\ell}_i)
\right\}_{i\in\mathcal I_{\mathrm A}},
$ and~$
\mathcal X_{\mathrm N}
=
\left\{
(\mathbf c_j,\boldsymbol{\ell}_j)
\right\}_{j\in\mathcal I_{\mathrm N}}.
$
Let~$
\mathcal L_{\mathrm A}=\{
\boldsymbol{\ell}_i
\}_{i\in\mathcal I_{\mathrm A}},
$ and~$
\mathcal L_{\mathrm N}
=
\{
\boldsymbol{\ell}_j
\}_{j\in\mathcal I_{\mathrm N}}
$
denote the corresponding collections of patch latents.
Following the factorization
\begin{equation}
p(\mathcal L_{\mathrm A},\mathcal L_{\mathrm N})
=
p(\mathcal L_{\mathrm A})
p(\mathcal L_{\mathrm N}\mid\mathcal L_{\mathrm A}),
\end{equation}
we process anchor patches independently in parallel, and then process non-anchor patches in parallel conditioned on anchor latents, as described next.

\subsubsection{Cross-Point Transformer (CPT)}

\begin{figure*}[t]
    \centering
    \includegraphics[width=\textwidth]{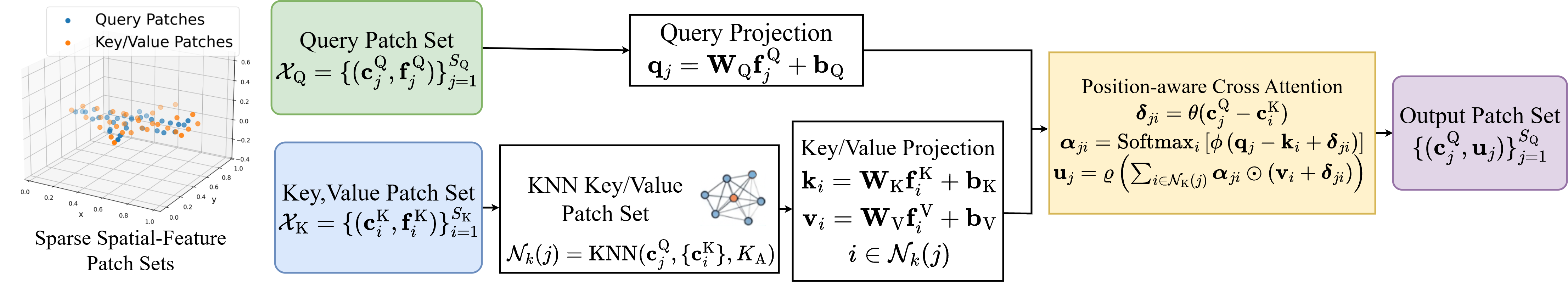}
    \caption{Overview of the CPT. It takes a query patch set $\mathcal X_{\mathrm Q} = \big\{ (\mathbf c_j^{\mathrm Q},\mathbf f_j^{\mathrm Q}) \big\}$ and a key/value patch set $\mathcal X_{\mathrm K} = \big\{ (\mathbf c_i^{\mathrm K},\mathbf f_i^{\mathrm K}) \big\}$ as inputs, where $\mathbf f_j^{\mathrm Q}$
    and $\mathbf f_i^{\mathrm K}$ denote the patch features at the corresponding CPT stage. For each query patch, its $K_{\mathrm A}$ nearest key/value patches are selected by KNN over patch centroids, and position-aware cross attention aggregates their projected features to produce the contextualized features.}
    \label{fig:system_architecture}
\end{figure*}

We use a CPT architecture, illustrated in Fig.~\ref{fig:system_architecture}, repeatedly in the next modules.
The CPT takes two sets of spatial-feature pairs,
namely, a query set $
\mathcal X_{\mathrm Q}
=
\big\{
(\mathbf c_j^{\mathrm Q},\mathbf f_j^{\mathrm Q})
\big\}_{j=1}^{S_{\mathrm Q}}
$ and a key/value set~$\mathcal X_{\mathrm K}
=
\left\{
(\mathbf c_i^{\mathrm K},\mathbf f_i^{\mathrm K})
\right\}_{i=1}^{S_{\mathrm K}}
$, as inputs.
For each query patch $j$, we select its $K_{\mathrm A}$ nearest
key/value patches according to their centroid distances, indexed by~
$
\mathcal N_{\mathrm K}(j)
=
\operatorname{KNN}
\left(
\mathbf c_j^{\mathrm Q},
\left\{
\mathbf c_i^{\mathrm K}
\right\}_{i=1}^{S_{\mathrm K}},
K_{\mathrm A}
\right).
$
The query, key, and value vectors are obtained through learned linear projections
\begin{equation}
\mathbf q_j = \mathbf W_{\mathrm Q}\mathbf f_j^{\mathrm Q}+\mathbf b_{\mathrm Q},\quad
\mathbf k_i = \mathbf W_{\mathrm K}\mathbf f_i^{\mathrm K}+\mathbf b_{\mathrm K},\quad
\mathbf v_i = \mathbf W_{\mathrm V}\mathbf f_i^{\mathrm K}+\mathbf b_{\mathrm V},
\end{equation}
respectively, for $i\in\mathcal N_{\mathrm K}(j)$. Here,
$\mathbf W_{\mathrm Q}\in\mathbb R^{D\times d_{\mathrm Q}}$ and
$\mathbf W_{\mathrm K},\mathbf W_{\mathrm V}
\in\mathbb R^{D\times d_{\mathrm K}}$
are learned projection matrices, while
$\mathbf b_{\mathrm Q},\mathbf b_{\mathrm K},\mathbf b_{\mathrm V}
\in\mathbb R^{D}$
are the corresponding bias vectors.
The relative position embedding between query patch~$j$
and its $i$-th neighboring key/value patch is
$
\boldsymbol\delta_{ji} = \theta(\mathbf c_i^{\mathrm K}-\mathbf c_j^{\mathrm Q}),
$
where
$\theta \colon \mathbb R^3\rightarrow\mathbb R^D$
is a learned positional embedding function.
The context-enhanced query feature is then computed as
\begin{align}
\mathbf u_j
&=
\varrho\Bigg(
\sum_{i\in\mathcal N_{\mathrm K}(j)}
\boldsymbol\alpha_{ji}\odot
(\mathbf v_i+\boldsymbol\delta_{ji})
\Bigg),\\
\boldsymbol\alpha_{ji} &= \operatorname{Softmax}_{i} \left[ \phi \left( \mathbf q_j-\mathbf k_i+\boldsymbol\delta_{ji} \right) \right],
\end{align}
where 
    $\varrho$ is a learned fully connected layer, 
$\phi\colon \mathbb R^D\rightarrow\mathbb R^D$
produces vector-valued attention logits, and
$\odot$ denotes element-wise multiplication.
Accordingly,
$\boldsymbol\alpha_{ji}\in\mathbb R^D$
contains attention weights, with the softmax applied over
$i\in\mathcal N_{\mathrm K}(j)$ independently for each feature
dimension.
Finally, the output of the CPT is given by
\begin{equation}
\mathcal F_{\mathrm{CPT}}
\left(
\mathcal X_{\mathrm Q},
\mathcal X_{\mathrm K}
\right)
=
\big\{
(\mathbf c_j^{\mathrm Q},\mathbf u_j)
\big\}_{j=1}^{S_{\mathrm Q}}.
\label{eq:cpt_output}
\end{equation}

The CPT architecture is instantiated in the spatial context encoder/decoder and conditional entropy model, with different query and key/value features as specified next.

\subsubsection{Non-Anchor Spatial Context Encoder and Decoder}
\label{sec:anchor_transmission}

The spatial context encoder takes $\mathcal X_{\mathrm N}$ and $\mathcal X_{\mathrm A}$ as inputs and outputs the residual $\mathcal X_{\mathrm N}^{\mathrm{enc}}$ for the non-anchor patches. For each non-anchor patch $j\in\mathcal I_{\mathrm N}$, the neighboring anchor representations are first aggregated by the CPT:
\begin{equation}
\left\{
\left(
\mathbf c_j,
\mathbf u_j
\right)
\right\}_{j \in \mathcal{I}_{\rm N}}
=
\mathcal F_{\mathrm{CPT}}
\left(
\left\{\mathbf c_j , \mathbf f \right \}, 
\mathcal X_{\mathrm A} \right),
\end{equation}
where $\mathbf f$ is a learnable query vector shared between non-anchor patches.

The non-anchor latents are then
represented by the residuals
\begin{equation}
\mathbf h_j^{\mathrm{enc}}
=
\boldsymbol{\ell}_j-
\psi (\mathbf u_j),
\qquad j\in\mathcal I_{\mathrm N},
\end{equation}
where $\psi$ is a multilayer perceptron (MLP). 
We denote the resulting residual patch set by
$
\mathcal X_{\mathrm N}^{\mathrm{enc}}
=
\left\{
\left(
\mathbf c_j,
\mathbf h_j^{\mathrm{enc}}
\right)
\right\}_{j\in\mathcal I_{\mathrm N}}
$.

At the receiver, the reconstructed anchor latents together with
the recovered centroids form
$
\widehat{\mathcal X}_{\mathrm A}
=
\big\{
\big(
\widehat{\mathbf c}_i,
\widehat{\boldsymbol{\ell}}_i
\big)
\big\}_{i\in\mathcal I_{\mathrm A}} .
$
The JSCC decoder first reconstructs the transmitted non-anchor
residual as $\mathbf h_j^{\mathrm{dec}}$. The spatial context decoder
then combines the decoded residuals with the reconstructed anchor
representations $\widehat{\mathcal X}_{\mathrm A}$ to recover the non-anchor
latents. Within this module, the CPT uses each recovered non-anchor
centroid~$\widehat{\mathbf c}_j$ as the query position and
$\widehat{\mathcal X}_{\mathrm A}$ as the key/value set to obtain the
receiver-side feature $\widehat{\mathbf u}_j$. The final non-anchor latents are
reconstructed as
\begin{equation}
\widehat{\boldsymbol{\ell}}_j
=
\psi(\widehat{\mathbf u}_j)
+
\mathbf h_j^{\mathrm{dec}},
\qquad j\in\mathcal I_{\mathrm N}.
\end{equation}

The reconstructed anchor and non-anchor latents are then
passed to the patch feature decoder for local geometry
reconstruction.

\subsubsection{Conditional Entropy and Patch-Based Rate Adaptation}
\label{subsec:entropy_rate}

We model anchor latents using a learned Gaussian distribution, and non-anchor latents using a learned Gaussian distribution conditioned on neighboring anchor patches.
Specifically, the patch latent $\boldsymbol{\ell}_i$ is modeled as
$\mathcal{N}(\boldsymbol{\mu}_i,[\operatorname{diag}(\boldsymbol{\sigma}_i)]^2)$,
where $\operatorname{diag}(\boldsymbol{\sigma}_i)$ is a diagonal matrix whose diagonal entries are the elements of the standard deviation vector $\boldsymbol{\sigma}_i$.
For patch~$i$, let
$
\bar{\boldsymbol{\ell}}_i
=
Q(\boldsymbol{\ell}_i)
$
denote its quantized latent representation, where $Q(\cdot)$ is a deterministic element-wise scalar quantization. During training, scalar quantization is approximated by noising with uniform noise in~$(-1/2, 1/2)^D$;
element-wise rounding to the nearest integer is used during inference.

For anchor patches, the Gaussian parameters are shared, i.e., $\boldsymbol{\mu}_i = \boldsymbol{\mu}_{\mathrm{A}},  \boldsymbol{\sigma}_i = \boldsymbol{\sigma}_{\mathrm{A}}$ for $i \in \mathcal{I}_{\mathrm A}$,
where $\boldsymbol{\mu}_{\mathrm{A}}$ and $\boldsymbol{\sigma}_{\mathrm{A}}$ are learnable parameter vectors.
For a non-anchor patch $j \in \mathcal I_{\mathrm N}$, let
$\mathcal N_{\mathrm A}(j) \subset \mathcal I_{\mathrm A}$ denote the
indices of its $K_{\mathrm A}$ nearest anchor patches according to
centroid distance. 
We first obtain the anchor-conditioned feature
$\mathbf u^{\mathrm e}_j$ as
\begin{equation}
\big\{
(\mathbf c_j,\mathbf u^{\mathrm e}_j)
\big\}
=
\mathcal F_{\mathrm{CPT}}
\big(
\left\{
\mathbf c_j, \mathbf f^{\mathrm e}
\right\},
\left\{
\left(
\mathbf c_i,
\boldsymbol{\ell}_i
\right)
\right\}_{i\in\mathcal N_{\mathrm A}(j)}
\big),
\end{equation}
where $\mathbf f^{\mathrm e}$ is a learnable query vector shared between non-anchor patches. The Gaussian parameters $\boldsymbol{\mu}_j,\boldsymbol{\sigma}_j$ are then predicted by applying an MLP on $\mathbf u^{\mathrm e}_j$, with a softplus function to enforce nonnegativity of $\boldsymbol{\sigma}_j$.  

Let $\bar{\ell}_{i,d}$ denote the $d$-th element of
$\bar{\boldsymbol{\ell}}_i$, and let $\mu_{i,d}$ and $\sigma_{i,d}$
denote the corresponding predicted Gaussian mean and standard
deviation. The estimated coding cost of patch $i$ is
\begin{equation}
R_i
=
-\sum_{d=1}^{D}
\log_2
P\!
\left(
\bar{\ell}_{i,d}
\mid
\mu_{i,d},
\sigma_{i,d}
\right) \quad \text{bits},
\label{eq:patch_coding_cost}
\end{equation}
where the probability mass is obtained by integrating the predicted
Gaussian distribution over the corresponding unit-width quantization
interval.
Following entropy-guided variable-rate JSCC~\cite{ntscc}, we
use $R_i$ to determine the channel resources
allocated to patch~$i$. Specifically, we define the continuous
rate surrogate of patch $i$ as
$\widetilde{k}_i = \alpha R_i$
, where $\alpha>0$ is a scaling factor.
The allocated length $k_i$ for patch $i$ is finally selected as the value in $\mathcal{K}=\{\kappa_1,\ldots,\kappa_M\}$ closest to $\widetilde{k}_i$.

\subsubsection{Rate-Adaptive Latent Transmission}

Guided by the rate allocation, the rate-adaptive JSCC encoder maps the representation of each patch\textemdash an anchor latent $\boldsymbol{\ell}_i$ or a non-anchor residual $\mathbf{h}^{\mathrm{enc}}_i$\textemdash to a complex-valued vector
$\mathbf{z}_i\in\mathbb{C}^{k_i}$
with~$\left\lVert\mathbf{z}_i\right\rVert_2^2=k_iP$.
The selected rate level is represented by a learned embedding vector. If the channel SNR is known by the transmitter and receiver, we also represent it using a learned embedding vector.
These embedding vectors are concatenated with the anchor latents $\boldsymbol{\ell}_i$ and non-anchor residual $\mathbf h_j^{\mathrm{enc}}$ at input of the rate-adaptive encoder, and with the received latents at the input of the rate-adaptive decoder.

\subsubsection{Training JSCC}
\label{subsec:training}

The overall objective function to train JSCC is as in~\eqref{RD_lagrangian}, except that the CBR is replaced by the surrogate $\frac{1}{S}
\sum_{i=1}^{S}\alpha R_i$ for differentiability. 
Training proceeds in two stages. In Stage 1, the feature encoder and
decoder are pretrained as noise-free autoencoder by feeding the patch latent representations
$\{ \boldsymbol{\ell}_i \}^S_{i=1}$ to the feature decoder, following the bypass in
Fig.~\ref{fig:system_overview}. 
In Stage 2, we freeze the feature encoder and decoder, and jointly train the remaining modules.

\subsection{Digital Transmission of Centroids and Rate Allocation}
\label{subsec:lossless_cu}

Within the anchor and non-anchor groups, the patch centroids $\mathbf{c}_i$ and the rate allocation indicators are sorted in the same order, and transmitted via digital links. 
Representing each coordinate of each centroid using $b$ bits entails $3Sb$ bits in total. The number of possible rate allocation count vectors is $\binom{S_{\mathrm A}+M-1}{M-1}$ for the anchor patches and $\binom{S-S_{\mathrm A}+M-1}{M-1}$ for the non-anchor patches. The total payload is therefore
\begin{align}
    &B
    = \notag \\
    &~
    3Sb
    +
    \left\lceil\log_2\!
        \binom{S_{\mathrm A}+M-1}{M-1}
    + \log_2\!
        \binom{S-S_{\mathrm A}+M-1}{M-1}
    \right\rceil
\end{align}
bits.

Let
$\epsilon(B,n,\mathrm{SNR})$ denote the block error
probability of transmitting $B$ bits over the complex-valued AWGN channel with blocklength $n$, characterized in~\cite{polyanskiy2010channel}.
Under automatic repeat request with independent retransmissions, perfect error detection, and negligible feedback overhead, the expected number of channel uses is 
$
    {n}/
    ({1-\epsilon(B,n,\mathrm{SNR})}),
$
which we minimize over $n$.

Finally, with analog and digital transmission combined, the 
CBR is computed as
\begin{equation}
    \mathrm{CBR}(\mathcal P)
    =
    \frac{\min_n \{{n}/({1-\epsilon(B,n,\mathrm{SNR})})\}
        +
        \sum_{i=1}^{S}k_i
    }{3N}.
\end{equation}
Note that it varies with the SNR.

\section{Numerical Experiments}

\subsection{Experimental Setup}
\label{sebsec:4a}
We evaluate the proposed method on ModelNet40~\cite{wu15_modelnet}.
Following the preprocessing procedure in~\cite{semcom}, each point cloud contains $N=8192$ points. The patch latent dimension is set to $D=64$.
The supported channel-use levels are~$\mathcal{K}=\{2,4, 6, 8,12,16,20,24,28, 32\}$. Different CBR operating points are obtained by varying the number of patches $S$ and the scaling factor $\alpha$.
Throughout, we set the rate-distortion trade-off coefficient to $\lambda=10^{-4}$, oversampling factor $\rho = 2$, and the number of neighboring anchor patches~$K_{\rm A}=3$. We designate the first half of the patches as anchors, i.e.,~$S_{\rm A}=S/2$, and account for each centroid coordinate using~$b=16$ bits.

Both training stages last $40$ epochs with a batch size of $2$. During
Stage 2, the channel SNR is sampled from
$0$ to $10$~dB. We use Adam optimizer with a learning rate of $5\times 10^{-4}$ initially and reduced by a factor of 10 every 20 epochs.
We report the communication cost via CBR, and reconstruction quality via D1 and D2 PSNR values.

\subsection{Rate–Distortion Performance}

We compare our GeoContext-JSCC method with G-PCC~\cite{graziosi2020overview}, SEPT~\cite{wireless_point_cloud_transmission}, DPCC~\cite{zhang2022transformer}, and the method of Xie et al.~\cite{semcom}.

To vary the CBR, we vary the number of patches and the fixed patch rate for Xie et al.~\cite{semcom}, and the compression rate for G-PCC and DPCC.
For G-PCC and DPCC, we transmit the compressed bitstreams using rate-\(1/2\) polar coding and quadrature phase-shift keying (QPSK) modulation.
For SEPT, as in~\cite{semcom}, we plot the performance for ${\rm CBR} = 0.017$ only because higher-CBR configurations did not improve D1 or D2 PSNR in our reproduction. This is consistent with the saturation behavior reported in~\cite{wireless_point_cloud_transmission}, possibly caused by the global max-pooling operation discarding fine-grained geometric information.

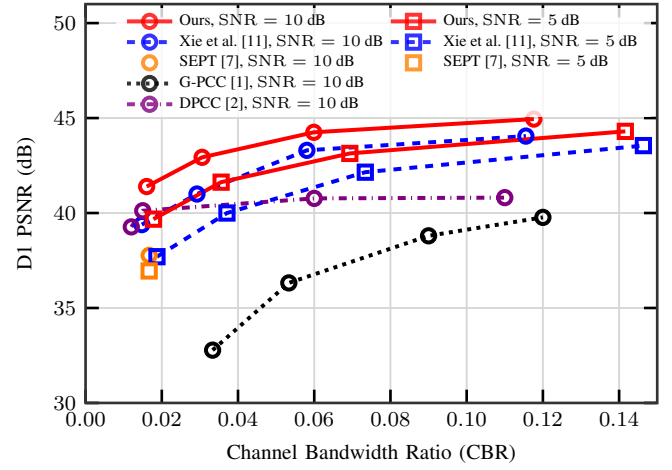
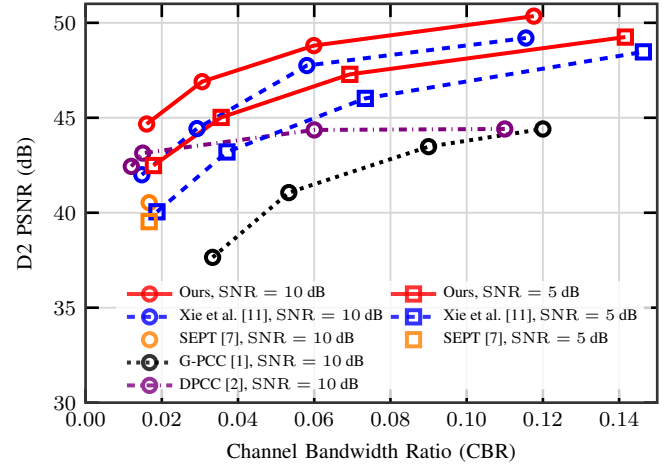
\begin{figure}[t!]
    \centering
    \subfloat[D1 PSNR versus CBR.\label{fig:rd_cpp_d1}]{
        \begin{tikzpicture}[font=\footnotesize]
    \begin{axis}[
      width=3.6in, height=2.7in,
      xmin=0, xmax=0.15,
      ymin=30, ymax=51,
      xtick={0,0.02,0.04,0.06,0.08,0.10,0.12,0.14},
      scaled ticks=false,
      xticklabel style={/pgf/number format/fixed,/pgf/number format/precision=2,
                        /pgf/number format/fixed zerofill},
      xlabel={Channel Bandwidth Ratio (CBR)},
      ylabel={D1 PSNR (dB)},
      axis line style={draw=black!80,line width=1pt},
      tick align=inside, tick pos=both,
      major tick length=5pt, tick style={black,line width=1pt},
      grid=major, grid style={draw=black!16,line width=0.7pt},
      legend columns=5,
      legend cell align={left},
      transpose legend,
      legend style={at={(.99,.99)}, anchor=north east,
        draw=none,fill=white,
        opacity=0.8,text opacity=1,
        font=\fontsize{6}{10}\selectfont,
        inner ysep=0pt,row sep=-1.8pt},
      unbounded coords=jump,
    ]
    \addplot[color=red, solid, line width=1.5pt, mark=o, mark size=2.5pt, mark options={solid,fill=none,line width=1.5pt}] coordinates {
      (0.01603,41.3939)
      (0.03058,42.9322)
      (0.05989,44.25)
      (0.11763,44.9469)
    };
    \addlegendentry{Ours, ${\rm SNR} = 10$~dB}

    \addplot[color=blue, dashed, line width=1.5pt, mark=o, mark size=2.5pt, mark options={solid,fill=none,line width=1.5pt}] coordinates {
      (0.0147666666667,39.39)
      (0.0292333333333,41)
      (0.0580666666667,43.3)
      (0.115533333333,44.05)
    };
    \addlegendentry{Xie et al.~\cite{semcom}, ${\rm SNR} = 10$~dB}

    \addplot[color=orange, only marks, mark=o, mark size=2.5pt, mark options={solid,fill=none,line width=1.5pt}] coordinates {
      (0.0166833333333,37.77)
    };
    \addlegendentry{SEPT~\cite{ota_learning_based}, ${\rm SNR} = 10$~dB}

    \addplot[color=black, dotted, line width=1.5pt, mark=o, mark size=2.5pt, mark options={solid,fill=none,line width=1.5pt}] coordinates {
      (0.0333333333333,32.78)
      (0.0533333333333,36.32)
      (0.09,38.8)
      (0.12,39.77)
    };
    \addlegendentry{G-PCC~\cite{graziosi2020overview}, ${\rm SNR} = 10$~dB}
    \addplot[color=violet, dashdotted, dash phase=0pt, line width=1.5pt, mark=o, mark size=2.5pt, mark options={solid,fill=none,line width=1.5pt}] coordinates {
      (0.012,39.27)
      (0.015,40.12)
      (0.06,40.77)
      (0.11,40.81)
    };
    \addlegendentry{DPCC~\cite{zhang2022transformer}, ${\rm SNR} = 10$~dB}

    \addplot[color=red, solid, line width=1.5pt, mark=square, mark size=2.5pt, mark options={solid,fill=none,line width=1.5pt}] coordinates {
      (0.0178,39.6775)
      (0.03557,41.6248)
      (0.06934,43.1377)
      (0.14163,44.299)
    };
    \addlegendentry{Ours, ${\rm SNR} = 5$~dB}

    \addplot[color=blue, dashed, line width=1.5pt, mark=square, mark size=2.5pt, mark options={solid,fill=none,line width=1.5pt}] coordinates {
      (0.0187333333333,37.7)
      (0.0371,40)
      (0.0734,42.15)
      (0.146366666667,43.55)
    };
    \addlegendentry{Xie et al.~\cite{semcom}, ${\rm SNR} = 5$~dB}

    \addplot[color=orange, only marks, mark=square, mark size=2.5pt, mark options={solid,fill=none,line width=1.5pt}] coordinates {
      (0.0166833333333,36.95)
    };
    \addlegendentry{SEPT~\cite{ota_learning_based}, ${\rm SNR} = 5$~dB}

    \addlegendimage{empty legend}
    \addlegendentry{}
    \addlegendimage{empty legend}
    \addlegendentry{}
    \end{axis}
\end{tikzpicture}
    }
    
    \subfloat[D2 PSNR versus CBR.\label{fig:rd_cpp_d2}]{
        \begin{tikzpicture}[font=\footnotesize]
    \begin{axis}[
      width=3.6in, height=2.7in,
      xmin=0, xmax=0.15,
      ymin=30, ymax=51,
      xtick={0,0.02,0.04,0.06,0.08,0.10,0.12,0.14},
      scaled ticks=false,
      xticklabel style={/pgf/number format/fixed,/pgf/number format/precision=2,
                        /pgf/number format/fixed zerofill},
      xlabel={Channel Bandwidth Ratio (CBR)},
      ylabel={D2 PSNR (dB)},
      axis line style={draw=black!80,line width=1pt},
      tick align=inside, tick pos=both,
      major tick length=5pt, tick style={black,line width=1pt},
      grid=major, grid style={draw=black!16,line width=0.7pt},
      legend columns=5,
      legend cell align={left},
      transpose legend,
      legend style={at={(.99,.01)}, anchor=south east,
        draw=none,fill=white,
        opacity=0.8,text opacity=1,
        font=\fontsize{6}{10}\selectfont,
        inner ysep=0pt,row sep=-1pt},
      unbounded coords=jump,
    ]
\addplot[color=red, solid, line width=1.5pt, mark=o, mark size=2.5pt, mark options={solid,fill=none,line width=1.5pt}] coordinates {
  (0.01603,44.6717)
  (0.03058,46.9025)
  (0.05989,48.8)
  (0.11763,50.3522)
};
\addlegendentry{Ours, ${\rm SNR} = 10$~dB}

\addplot[color=blue, dashed, line width=1.5pt, mark=o, mark size=2.5pt, mark options={solid,fill=none,line width=1.5pt}] coordinates {
  (0.0147666666667,42)
  (0.0292333333333,44.43)
  (0.0580666666667,47.75)
  (0.115533333333,49.2)
};
\addlegendentry{Xie et al.~\cite{semcom}, ${\rm SNR} = 10$~dB}

\addplot[color=orange, only marks, mark=o, mark size=2.5pt, mark options={solid,fill=none,line width=1.5pt}] coordinates {
  (0.0166833333333,40.53)
};
\addlegendentry{SEPT~\cite{ota_learning_based}, ${\rm SNR} = 10$~dB}

\addplot[color=black, dotted, line width=1.5pt, mark=o, mark size=2.5pt, mark options={solid,fill=none,line width=1.5pt}] coordinates {
  (0.0333333333333,37.65)
  (0.0533333333333,41.06)
  (0.09,43.48)
  (0.12,44.41)
};
\addlegendentry{G-PCC~\cite{graziosi2020overview}, ${\rm SNR} = 10$~dB}

\addplot[color=violet, dashdotted, dash phase=0pt, line width=1.5pt, mark=o, mark size=2.5pt, mark options={solid,fill=none,line width=1.5pt}] coordinates {
  (0.012,42.44)
  (0.015,43.14)
  (0.06,44.36)
  (0.11,44.41)
};
\addlegendentry{DPCC~\cite{zhang2022transformer}, ${\rm SNR} = 10$~dB}

\addplot[color=red, solid, line width=1.5pt, mark=square, mark size=2.5pt, mark options={solid,fill=none,line width=1.5pt}] coordinates {
  (0.0178,42.4847)
  (0.03557,45.0182)
  (0.06934,47.2861)
  (0.14163,49.2498)
};
\addlegendentry{Ours, ${\rm SNR} = 5$~dB}

\addplot[color=blue, dashed, line width=1.5pt, mark=square, mark size=2.5pt, mark options={solid,fill=none,line width=1.5pt}] coordinates {
  (0.0187333333333,40.05)
  (0.0371,43.2)
  (0.0734,46.02)
  (0.146366666667,48.47)
};
\addlegendentry{Xie et al.~\cite{semcom}, ${\rm SNR} = 5$~dB}

\addplot[color=orange, only marks, mark=square, mark size=2.5pt, mark options={solid,fill=none,line width=1.5pt}] coordinates {
  (0.0166833333333,39.53)
};
\addlegendentry{SEPT~\cite{ota_learning_based}, ${\rm SNR} = 5$~dB}

    \addlegendimage{empty legend}
    \addlegendentry{}
    \addlegendimage{empty legend}
    \addlegendentry{}
    \end{axis}
\end{tikzpicture}
    }
    \caption{Rate--distortion performance under
    different channel-use budgets.}
    \vspace{-6mm}
    \label{fig:rd_cpp}
\end{figure}

In Fig.~\ref{fig:rd_cpp}(a) and Fig.~\ref{fig:rd_cpp}(b), we report the D1 and D2 rate--distortion performance, respectively, achieved by each method under similar CBR values. Our GeoContext-JSCC method outperforms all other considered methods across the tested CBR range and ${\rm SNR} \in \{5, 10\}$~dB. The gap to Xie et al.~\cite{semcom} at ${\rm SNR} = 10$~dB 
is $0.55$--$1.65$~dB in D1 PSNR and
$1.05$--$2.67$~dB in D2 PSNR. The D1 and D2 PSNR gain is more pronounced at lower CBRs, indicating that GeoContext-JSCC is effective when communication resources are limited.

\subsection{Ablation Studies}
\label{subsec:ablation}

We further conduct ablation studies to evaluate the contribution of rate allocation and anchor-based spatial context.
All below use the same latent dimension, training schedule, channel settings, and reconstruction backbone. We set $S=64$ patches, corresponding to $K=256$ points per patch.

\subsubsection{Effect of Variable-Rate Allocation}
We consider a variant without variable rate allocation, in which all patches are assigned the same symbol-sequence length as $k = 3$ complex values. The resulting average CBR is comparable to that of our model with rate allocation.

\subsubsection{Effect of Spatial Context}
We consider a variant without the spatial context encoder and decoder, in which all patches are treated independently as anchors.

\begin{figure}[t!]
    \centering
\begin{tikzpicture}[font=\footnotesize]
\begin{axis}[
  width=3.6in, height=2.8in,
  xmin=0, xmax=10, ymin=30, ymax=51,
  scaled ticks=false,
  xlabel={Channel SNR (dB)}, ylabel={PSNR (dB)},
  axis line style={draw=black!80,line width=1pt},
  tick align=inside, tick pos=both,
  major tick length=5pt, tick style={black,line width=1pt},
  grid=major, grid style={draw=black!16,line width=0.7pt},
  legend cell align={left},
  legend style={at={(.99,.01)}, anchor=south east,
    draw=none,fill=white,
    opacity=0.8,text opacity=1,
    inner ysep=0pt,row sep=-1.8pt},
  clip=true, clip mode=individual,
]

\addplot[color=red, solid, line width=1.5pt] coordinates {
  (0,42.3631)
  (2,42.9264)
  (4,43.4208)
  (5,43.6219)
  (10,44.3016)
};
\addlegendentry{Ours}

\addplot[forget plot, color=red, solid, line width=1.5pt] coordinates {
  (0,46.0681)
  (2,46.6103)
  (4,47.5000)
  (5,47.6105)
  (10,48.6623)
};

\addplot[color=blue, loosely dashed, line width=1.5pt] coordinates {
  (0,41.5491)
  (2,42.0733)
  (4,42.8441)
  (5,43.0521)
  (10,43.8675)
};
\addlegendentry{without spatial context coding}

\addplot[forget plot, color=blue, loosely dashed, line width=1.5pt] coordinates {
  (0,45.1221)
  (2,45.9489)
  (4,46.6231)
  (5,46.7988)
  (10,47.9742)
};

\addplot[color=orange, dashdotted, line width=1.5pt] coordinates {
  (0,41.5651)
  (2,42.0779)
  (4,42.5489)
  (5,42.7546)
  (10,43.4152)
};
\addlegendentry{without variable rate allocation}

\addplot[forget plot, color=orange, dashdotted, line width=1.5pt] coordinates {
  (0,44.9229)
  (2,45.5773)
  (4,46.1988)
  (5,46.4782)
  (10,47.4246)
};

\addplot[color=black, dotted, line width=1.5pt] coordinates {
  (0,17.98)
  (2,26.39)
  (4,40.63)
  (5,40.8)
  (10,40.81)
};
\addlegendentry{DPCC~\cite{zhang2022transformer}}

\addplot[forget plot, color=black, dotted, line width=1.5pt] coordinates {
  (0,17.98)
  (2,27.72)
  (4,44.21)
  (5,44.37)
  (10,44.41)
};

\addplot[color=teal!70!black, densely dashed, line width=1.5pt] coordinates {
  (0,17.98)
  (2,24.09)
  (4,41.10)
  (5,41.30)
  (10,41.30)
};
\addlegendentry{G-PCC~\cite{graziosi2020overview}}

\addplot[forget plot, color=teal!70!black, densely dashed, line width=1.5pt] coordinates {
  (0,17.98)
  (2,25.17)
  (4,44.98)
  (5,45.63)
  (10,45.63)
};

\draw[black, line width=0.7pt]
  (axis cs:5,41.8) ellipse[x radius=4pt, y radius=16pt];
\draw[black, line width=0.7pt]
  (axis cs:4.5,46) ellipse[x radius=4pt, y radius=16pt];

\node (d1label) at (axis cs:6.5,38.5) {D1 PSNR};
\draw[->, >=stealth, black, line width=0.7pt]
  (d1label.west) -- (axis cs:5,39.6);

\node (d2label) at (axis cs:3,49) {D2 PSNR};
\draw[->, >=stealth, black, line width=0.7pt]
  (d2label.east) -- (axis cs:4.5,48.2);


\end{axis}
\end{tikzpicture}
    \vspace{-.5cm}
    \caption{Ablation results for variants without spatial context coding or variable rate allocation under different channel SNRs.}
    \label{fig:ablation_snr}
\end{figure}
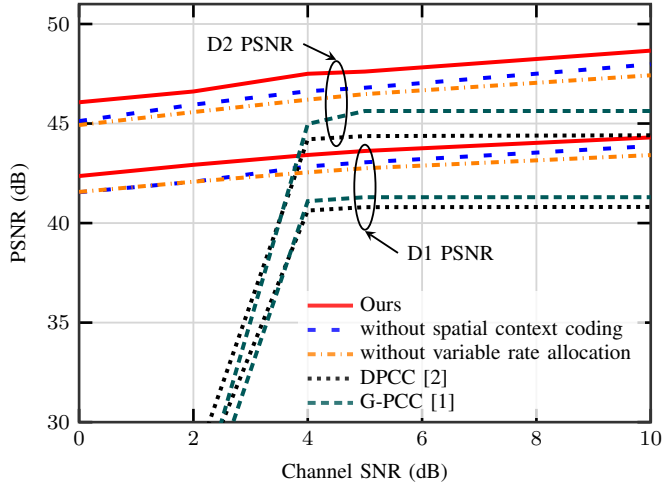

In Fig.~\ref{fig:ablation_snr}, we show the ablation results under different channel SNRs.
Because of the variation of the block length $\min_n \{{n}/({1-\epsilon(B,n,\mathrm{SNR})})\}$ across SNR values, the CBR decreases from approximately $0.142$ at $0$~dB to $0.046$ at $10$~dB as the channel SNR increases.
The full model with rate allocation and spatial context consistently outperforms the variants without either of these components in both D1 and D2 PSNR. Replacing the entropy-guided variable-rate allocation with a fixed sequence length results in a clear performance loss, showing that patch-adaptive rate allocation improves the utilization of the available channel budget. Similarly, removing anchor-based spatial context degrades reconstruction quality across the tested SNRs, demonstrating the overall benefit of anchor-conditioned entropy estimation, encoding, and decoding. These results indicate that adaptive rate allocation and anchor-conditioned context provide complementary gains. For reference, we also report the DPCC and G-PCC baseline at a fixed CBR of around $0.173$.
Three baselines exhibit a clear cliff effect at a certain SNR, and leveling effect as the SNR increases, whereas our GeoContext-JSCC method degrades gracefully and maintains higher reconstruction quality.

\vspace{-1mm}

\subsection{Visual Comparison}
\label{subsec:visualization}

\begin{figure}[t]
    \centering
    \subfloat[Airplane\label{fig:vis_plane}]{
        \includegraphics[width=\linewidth]{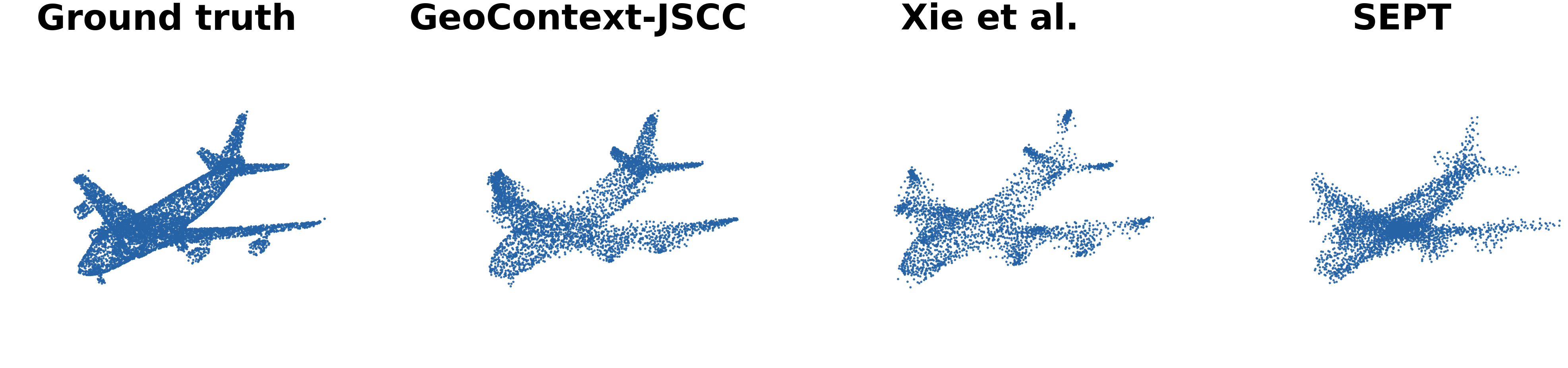}
    }

    \vspace{0mm}

    \subfloat[Bed\label{fig:vis_bed}]{
        \includegraphics[width=\linewidth]{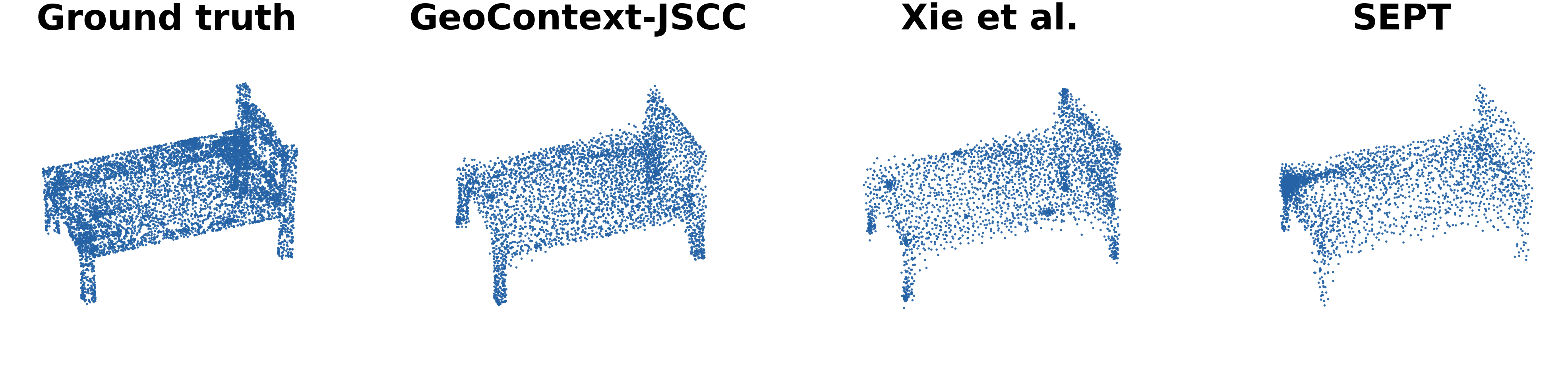}
    }

    \vspace{0mm}

    \caption{Visual comparison of point cloud reconstruction
    on ModelNet40.
    From left to right: ground truth, GeoContext-JSCC,
    Xie et al.~\cite{semcom}, and
    SEPT~\cite{wireless_point_cloud_transmission}.}
    \label{fig:visual_comparison}
    \vspace{-5mm}
\end{figure}

We present a visual comparison of reconstructed point clouds obtained by considered JSCC methods in Fig.~\ref{fig:visual_comparison}, with ${\rm CBR} \approx 0.017$ and SNR~$=10$~dB. For GeoContext-JSCC, we set $S=16$ and $K=1024$. We randomly select test samples 
and reconstruct them using GeoContext-JSCC, Xie et al.~\cite{semcom}, and SEPT~\cite{wireless_point_cloud_transmission}.
This represents a particularly low-CBR regime, where only limited channel resources are available. 
As can be seen, GeoContext-JSCC better preserves the overall object shapes and exhibits a more uniform spatial distribution of reconstructed points, whereas the competing methods show more noticeable geometric distortions and irregular point concentrations.

\section{Conclusion}

We propose GeoContext-JSCC, a contextual variable-rate JSCC framework for wireless point cloud transmission.
The framework exploits local dependencies among neighboring patches through anchor-derived context to estimate the conditional entropy of patch latents and determine per-patch transmission lengths.
The anchor-based spatial context allows our model to exploit inter-patch dependencies. Experiments on ModelNet40 over AWGN channels demonstrate improved rate-distortion performance over the considered conventional digital transmission and learning-based JSCC baselines. 
Ablation results confirm the contributions of both variable-rate allocation and anchor-conditioned entropy modeling.

\bibliographystyle{IEEEtran}
\bibliography{references}

\end{document}